\documentstyle[12pt]{article}
\renewcommand{\a}{\alpha}
\renewcommand{\b}{\beta}
\renewcommand{\d}{\delta}

\newcommand{\g}{\gamma}

\renewcommand{\l}{\lambda}
\newcommand{\m}{\mu}
\newcommand{\n}{\nu}

\renewcommand{\o}{\omega}

\newcommand{\r}{\rho}
\newcommand{\s}{\sigma}

\newcommand{\th}{\theta}
\newcommand{\vf}{{\varphi}}

\newcommand{\G}{\Gamma}

\newcommand{\ad}{{\dot{\alpha}}}
\newcommand{\dd}{{\,D\!D\,}}
\newcommand{\ddb}{{\,{\bar D}\!{\bar D}\,}}
\newcommand{\Db}{{\bar D}}
\newcommand{\Ab}{{\bar A}}
\newcommand{\cc}{{{\rm c.c.}}}
\newcommand{\thb}{{\bar\theta}}

\newcommand{\real}{{{\rm I} \kern -.19em {\rm R}}}
\newcommand{\tr}{{\rm {Tr} \,}}
\newcommand{\half}{\frac 1 2}
\newcommand{\pa}{\partial}

\newcommand{\intd}{{\int d^4 \! x \,\,}}

\newcommand{\WWb}{{\bar{\cal W}}}
\newcommand{\ie}{{{\em i.e.},\ }}
\newcommand{\eg}{{{\em e.g.},\ }}

\newcommand{\NN}{{\cal N}}

\newcommand{\RR}{{\cal R}}
\newcommand{\SS}{{\cal S}}

\newcommand{\WW}{{\cal W}}

\newcommand{\lp}{\displaystyle{\bigl(}}
\newcommand{\rp}{\displaystyle{\bigr)}}
\newcommand{\Lp}{\displaystyle{\biggl(}}
\newcommand{\Rp}{\displaystyle{\biggr)}}

\newcommand{\lc}{\displaystyle{\bigl[}}
\newcommand{\rc}{\displaystyle{\bigr]}}
\newcommand{\Lc}{\displaystyle{\biggl[}}
\newcommand{\Rc}{\displaystyle{\biggr]}}

\newcommand{\la}{\displaystyle{\bigl\{}}
\newcommand{\ra}{\displaystyle{\bigr\}}}

\newcommand{\be}{\begin{equation}}
\newcommand{\ee}{\end{equation}}
\newcommand{\een}[1]{\label{#1}\end{equation}}
\newcommand{\beq}{\begin{eqnarray}}
\newcommand{\eeq}{\end{eqnarray}}
\newcommand{\eeqn}[1]{\label{#1}\end{eqnarray}}
\newcommand{\ba}{\begin{array}}
\newcommand{\ea}{\end{array}}
\newcommand{\brcl}{\begin{equation}\begin{array}{rcl}}
\newcommand{\ercl}{\end{array}\end{equation}}
\newcommand{\ercln}[1]{\end{array}\label{#1}\end{equation}}
\newcommand{\nn}{\nonumber}
\renewcommand{\=}{&=&} 

\newcommand{\equ}[1]{(\ref{#1})}
\newcommand{\journal}[4]{{\em #1~}#2,\,#3\,(19#4);}
\newcommand{\hpa}{\journal {Helv. Phys. Acta}}
\newcommand{\ijmp}{\journal {Int. J. Mod. Phys.}}
\newcommand{\pr}{\journal {Phys. Rev. D}}

\newcommand{\cmp}{\journal {Comm. Math. Phys.}}

\newcommand{\np}{\journal {Nucl. Phys.}}
\newcommand{\pl}{\journal {Phys. Lett.}}

\newcommand{\ptps}{\journal {Progr. Theor. Phys. Suppl. }}

\makeatletter
\@addtoreset{equation}{section}
\makeatother

\begin{document}
\setlength{\baselineskip}{3.0ex}
\thispagestyle{empty}
\noindent{\normalsize October 1995} \hfill {\normalsize NEIP-95-012}\\
\noindent\null \hfill {\normalsize hep-th/9510078}\\
\vspace{2.65cm}

\begin{center}
{\Large\bf ALL-ORDERS FINITE}\\[4mm]
{\Large\bf $N=$1 SUPER-YANG-MILLS THEORIES\footnote{Talk
given at SUSY 95, Paris, June
15-19, 1995. Work supported in part by the Swiss National
Science Foundation.}}

\vspace*{6.0ex}

{\normalsize Claudio LUCCHESI}

\vspace*{1.5ex}

{\normalsize\it Institut de Physique, Universit\'e de Neuch\^atel}\\
{\normalsize\it 1 rue Br\'eguet, CH -- 2000 Neuch\^atel (Switzerland)}\\
{\normalsize E-mail: lucchesi@iph.unine.ch}\\

\vspace*{6cm}

{\bf Abstract}

\end{center}
\vspace*{-1.0ex}
I present a criterion for all-order finiteness
in $N=1$ SYM theories. The structure of the
supercurrent anomaly, the Callan-Symanzik
equation and the supersymmetric
non-renormalization theorem for chiral
anomalies are the essential ingredients
of the proof.

\setcounter{footnote}{0}
\setcounter{page}{0}
\newpage
\section{Introduction}

I shall report in this talk about a criterion for all-order
finiteness in $N=1$ supersymmetric grand
unified theories (SGUTs). By all-order finiteness, I mean
the vanishing of all $\b$-functions, both
gauge and Yukawa, at every order of perturbation theory.
This finiteness criterion \cite{lucchesipiguetsiboldHPA}
has been on the market for some time,
and it is now being applied to phenomenological models.
Following a talk in that direction by G.
Zoupanos \cite{zoupanosetc}, I would like to
present in some details the criterion itself, as well
as some steps of its derivation.

Before I start, let me mention that there exist
related (and slightly different) approaches to
all-order finiteness in $N=1$ SYM
\cite{ermushev,chinese}. Due to  lack of time,
I shall not comment on these.

\section{$N=1$ Super-Yang-Mills Theory}

We consider an $N=1$ SYM theory with a
simple gauge group $G$.
The real gauge and chiral matter
superfields are resp. denoted by $\phi$
and\footnote{For matter superfields, we
use a compact indices convention: $R$ denotes
both the field and its representation. We also define $r\equiv
(R,\rho)$, where $\rho$ denotes the field components
within a given representation $R$.} $A^R$.
The gauge-invariant superfield action writes (for
conventions, see
\cite{lucchesipiguetsiboldHPA,piguetsibold86}):
\beq
S^{\ \rm inv.}=&-&{1\over 128 g^2}\ \tr \intd \dd\ F^\a F_\a + {1\over
16}\ \intd \dd \ddb \sum_R \Ab_R \,e^{\phi_iT^i_R}\, A^R\nn\\
&+&{1\over 6}\ \Lp\intd \dd \l_{rst}\, A^rA^sA^t + \cc\Rp\ .
\label{invaction}
\eeq
The gauge-fixing of
the above action \cite{piguetsibold86} is out
of our purpose. Following the BRS quantization
procedure, we construct the vertex
functional as
$\G = S^{\ \rm inv.} + S^{\ \rm gauge\ fixing} +
S^{\ \rm Faddeev-Popov}
+$ loop corrections of order $\hbar^n$,
and define the quantum theory generating
functional to be the most general solution of a set of
constraints given by the gauge condition,
the equations of motion, the rigid and BRS
symmetries, etc. A subset of these constraints
is relevant to this presentation:

1. $\RR$-symmetry. Infinitesimally $\d_\RR\vf
=i\,( n_\vf+\th^\a\pa_{\th^\a}-\thb^\ad\pa_{\thb^\ad})\,\vf$
on a generic superfield $\vf=A,\Ab,\phi,c_+,{\bar c}_+,\dots$,
with $\RR$-weights
$n_A=-n_\Ab=-{\scriptstyle{2/3}}$, $n_{\vf=\phi,c_+,{\bar c}_+}
={\scriptstyle 0}$. The functional
$\RR$-Ward identity writes $\WW_\RR\G=0$.

2. Supersymmetry, expressed through the Ward identities
$\WW_\a \G=0$, $\WWb_\ad \G=0$.

3. BRS invariance, acting infinitesimally as
$s\,e^\phi=e^\phi c_+-{\bar c}_+e^\phi$,
$s\,A^r=-c_{+\,i}(T^i_R)^\r_{\ \s}A^{(R,\s)}$, and
$s\,c_+=-\half\la c_+;c_+\ra$.
BRS invariance is encoded in a (non-linear)
Ward identity, the Slavnov identity $\SS(\G)=0$,
which is satisfied provided there is no gauge
anomaly \cite{piguetsibold86,chiralanomaly}.

4. A possible set of rigid chiral symmetries:
$\d_aA^R=i\,e_{a\ S}^{\ R}A^S$,
$\d_a\Ab_R=-i\Ab_S\,e_{a\ R}^{\ S}$, generated by Hermitean
charges $e_a=e_a^\dagger$ ($\d_a\phi=\d_ac_+=0$).
The ``chiral" Ward identity\footnote{The Ward identities
for $\RR$-symmetry, supersymmetry,
BRS symmetry, as well as the chiral one, are taken to
hold up to soft, supersymmetric mass
terms.} $\WW_a\G=0$ is satisfied provided
$\l_{rsu}e_{a\ t}^{\ \,u} + {\rm cyclic\
permutations}(r,s,t) =0$.

\section{Supercurrent and Anomalies}

The Ward operators for supersymmetry,
 translations and
$\RR$-invariance obey a superPoincar\'e
algebra. As a consequence, there exists a superfield
Ward operator $\hat\WW=\WW_R-i\,\th^\a\WW_\a +
i\,\thb^\ad\WWb_\ad
-2(\th^\a\s^\m_{\a\ad}\thb^\ad)\,\WW_\m^T +\dots\ .$
The component N{\oe}ther currents associated to
$\RR$-symmetry, supersymmetry and
translation invariance form the
supercurrent \cite{ferrarazumino,piguetsiboldsc}
$V_\m(x,\th,\thb)=R_\m(x)-i\,\th^\a
Q_{\m\a}(x) + i\,\thb^\ad
{\bar Q}_{\m\ad}(x) - 2(\th^\a\s^\n_{\a\ad}\thb^\ad)\,
T_{\m\n}(x)+ \dots$, which satisfies the supertrace equation:
\be
{\hat\o}\G =\pa^\m V_\m+i\,(\dd{\bf S}-\ddb{\bf\bar S})\ .
\label{sutrace}
\ee
The chiral superfield ${\bf S}$ in the right
side is the supercurrent anomaly. In components, it
yields an Abelian chiral anomaly which breaks
the $R$-current divergence:
\be
\pa_\m R^\m = i\,{\hat\o}\G |_{\th=0} +
i\,(\dd{\bf S}-\ddb{\bf\bar S})\ ,
\label{anomrdiv}
\ee
as well as dilatation anomalies in the energy-momentum ``trace":
\be
\intd T^\m_{\ \m} = \WW^D\G -{1\over 3}\,\intd (\dd{\bf S}
+ \ddb{\bf\bar S})\ .
\label{anomemtrace}
\ee
Here $\WW^D$ is the Ward operator of dilatations:
 $\d_D\vf=( d_\vf+x^\m\pa_\m
+{\scriptstyle\half}\th^\a\pa_{\th^\a}
+{\scriptstyle\half}\thb^\ad\pa_{\thb^\ad})\ \vf$.

Our task is to relate the Abelian chiral anomaly in the
$\RR$-current divergence and the dilatation
anomalies in the energy-momentum ``trace" to the
Abelian anomalies associated to the (possible)
chiral symmetries $\WW_a$. The natural setting
for deriving such a relation is provided by the
Callan-Symanzik equation. We shall arrive at its
formulation by expanding the supercurrent
anomaly $\bf S$ in a basis of dimension 3,
BRS-invariant, chiral insertions $\{L_i$\} as:
\be
{\bf S} =\b_g\,L_g +\sum_{\l_{rst}} \b_{rst}\,L_{rst}
-\sum_{R,S}\g^S_{\ R}\,L^R_{\ S} + \dots\ ,
\label{sexp}
\ee
where the dots stand for insertions which are
not essential in the present context, and the $L_i$'s
are defined by:
\brcl
\intd (\dd L_g + \ddb {\bar L}_g )&\equiv& \pa_g\G\\[1.5mm]
\intd (\dd L_{rst} + \ddb {\bar L}_{rst} )&\equiv& \pa_{\l_{rst}}\G\\
\intd (\dd L^R_{\ S} + \ddb {\bar L}^R_{\ S} )
&\equiv&
\NN^R_{\ S}\G = \intd \Lp \dd A^R{\d\over\d A^S}
+\ddb {\bar A}_S{\d\over\d {\bar A}_R}\Rp\G\ .
\label{SL}
\ercl
Inserting into the energy-momentum ``trace" the
expansion for $\bf S$ and the forms of the
$L_i$'s, and relating the (broken) Ward identity of
dilatations to the scaling operator through the
dimensional analysis identity
$\WW^D\G=\sum_{\m_i}\m_i\,\pa_{\m_i}$, one arrives at:
\be
C\G\equiv \Lc  \sum_{\m_i}\m_i\pa_{\m_i} +\b_g\,\pa_g
+\sum_{\l_{rst}} \b_{rst}\,\pa_{\l_{rst}}
-\sum_{R,S}\g^S_{\ R}\,\NN^R_{\ S}
+ \dots \Rc \ \G =0\ .
\label{CS}
\ee
This is the Callan-Symazik equation, which
describes how dilation invariance is broken by the
$\b$-functions $\b_g$, $\b_{rst}$ associated
to the renormalization of the gauge, resp. Yukawa
couplings, and by the anomalous dimensions
$\g^S_{\ R}$.

Let us now perform a change of basis for the
counting operators $\NN^R_{\ S}$: $\la \NN^R_{\ S}
\ra \rightarrow
\la \NN_{0a}\equiv e_{0a\ S}^{\ \  R}\ \NN^R_{\ S}\ra
\bigoplus \la \NN_{1k}\ra$, where the
$e_{0a\ S}^{\ \ R}$ are charge matrices corresponding
to the center of the algebra of chiral
symmetries $\la\WW_a\ra$ (\ie $\lc \WW_{0a};\WW_b\rc =0,
\forall b$), and the new counting
operators annihilate the superpotential:
 $\NN_{0a} \lp \intd \dd \l_{rst}\, A^r A^s A^t  \rp =0$.

Next, one can show that the supercurrent
anomaly $\bf S$, as well as each of the insertions
$L_i$ of its expansion in the new basis
(omitting the unessential term $L_{1k}$),
can be written
as:
\brcl
{\bf S} \=\ddb\, ( r \, K_3^0 + \dots)\ ,\\
L_g \=\ddb \, ( {\scriptstyle{1\over 128 g^3}} + r_g )
\, K_3^0 + \dots\ ,\\
L_{rst} \=\ddb \, r_{rst} \, K_3^0 + \dots\ ,\\
L_{0a} \=\ddb \, r_{0a} \, K_3^0 + \dots\ ,
\ercl
where the dots stand for invariant currents and
``genuinely chiral" terms which cannot be written
as $\Db\Db(\dots)$.
Replacing these expressions into
${\bf S}= \b_g\,L_g +\sum_{\l_{rst}} \b_{rst}\,L_{rst}
-\sum_a\g_{0a}\,L_{0a} + \dots$, and
identifying the coefficients of the $K_3^0$-dependent
terms, yields the relation:
\be
r=\b_g\ \Lp {1\over 128\, g^3} + r_g\Rp +
\sum_{\l_{rst}}\b_{rst}\,r_{rst}
-\sum_a \g_{0a}\, r_{0a}\ .\label{rrr}
\ee

\section{Non-renormalization of Chiral Anomalies}

Specializing to the case under consideration,
the non-renormalization theorem for chiral
anomalies in $N=1$ SYM (see
\cite{lucchesipiguetsiboldHPA,piguetsiboldIJMPA})
tells us that $r$ and $r_{0a}$ in \equ{rrr} are
non-renormalized, \ie they are strictly of order $\hbar$.

$r$ is the coefficient of the Abelian anomaly
in the $\RR$-axial current, and the $r_{0a}$'s are
the coefficients of the Abelian anomalies of the
axial currents associated to the chiral
$\WW_a$-symmetries. $r$ and $r_{0a}$ are given
by their one-loop values
\cite{parkeswestjonesmezincescu,
piguetsiboldIJMPA,lucchesipiguetsiboldHPA}:
\be
r = {1\over 128\,g^3}\ \b^{(1)}_g
= {1\over 512\,(4\pi)^2}\ \Lp\sum_R T(R)-3C_2(G)\Rp\ ,\qquad
r_{0a} = -{1\over 256\,(4\pi)^2}\  \sum_R e_{a\ R}^{\ R} \,T(R)\ .
\label{rr0a}
\ee

Note: The proof of the non-renormalization
theorem uses the fact that the three-form $K^0_3$,
the supersymmetric Chern-Simons form, is
related through the supersymmetric descent
equations to the zero-form $K^3_0 = {1 \over 3}\ \tr\, c_+^3$,
the cubed ghost field
insertion. The non-renormalization theorem for
chiral vertices guarantees the finiteness of the
latter insertion.

\section{Criterion for all-order vanishing $\b$-functions}

{\it Criterion}: Consider an N=1 super-Yang-Mills
theory with simple gauge group. If\\
(i) there is no gauge anomaly,\\
(ii) the gauge $\b$-function vanishes
at one loop:
\be
\b_g^{(1)} =0\ ,
\ee
(iii) there exist solutions of the form
$\l_{rst} =\l_{rst}(g)$ to the conditions of
vanishing one-loop anomalous dimensions
\be
\g^{(1)\ R}_{\ \ \ \ \ S} =0\ ,
\ee
and (iv) this solution is {\it isolated} and
{\it non-degenerate} when considered as a solution of
the conditions of vanishing one-loop Yukawa $\b$-functions:
\be
\b^{(1)}_{rst} = \l_{rsu}\,\g^{(1)\ u}_{\ \ \ \ \ t} +
{\rm cyclic}\ {\rm permutations}(r,s,t) =0\ ,
\ee
then the theory depends on a single coupling
constant (the gauge coupling $g$)
with a $\b$-function which vanishes at all orders.

Let us give a sketch of the proof. With the
expressions for $r$ and $r_{0a}$ \equ{rr0a},
it follows from (ii) and (iii) that $r=0$, resp.\footnote{One
uses here a corollary to the main
non-renormalization theorem stated in Section 4:
the conditions $\g^{(1)\ R}_{\ \ \ \ \ S} =0$ are
compatible {\it iff} $r_{0a} =0$.} $r_{0a}=0$.
Then \equ{rrr} reduces to
\be
0 = \b_g\ \Lp {1\over 128\, g^3} + r_g\Rp
+\sum_{\l_{rst}}\b_{rst}\,r_{rst}
\ .\label{rrrred}
\ee
That the Yukawa couplings $\l_{rst}$ are proportional
to $g$ in the one-loop approximation as a
consequence of (iii) is clear from
\cite{parkeswestjonesmezincescu}:
\be
\g^{(1)\ r}_{\ \ \ \ \  s}={1\over (2\pi)^2}\
\Lp\bar\l^{ruv}\l_{suv} -
{1\over 16}g^2 \,C_2(R)\d^r_{\,s}\Rp\ .
\ee
At higher orders, $\l_{rst}=\l_{rst}(g)$ are formal
power series in $g$, and one needs to impose
for consistency that these functions satisfy the
reduction equations  \cite{OehmeZimmermann}:
\be
\b_{rst}=\b_g\, {\pa\l_{rst}\over \pa g}\ .
\label{zred}
\ee
One can show \cite{lucchesipiguetsiboldHPA}
that a solution to these equations exists at all
orders (and is unique) if the lowest-order
solution is isolated and non-degerate. At
one-loop, eq. \equ{zred} reduces to $\b^{(1)}_{rst}=0$;
this is just hypothesis (iv).

Next one replaces \equ{zred} into \equ{rrrred} and gets:
\be
0 = \b_g\ \Lp {1\over 128\, g^3} + r_g + \sum_{\l_{rst}}
{\pa\l_{rst}\over \pa g}\Rp\ .
\label{rrrredrepl}
\ee
The parenthesis being perturbatively invertible,
it follows that $\b_g=0$ at all orders, for the
unique remaining (independent) coupling of the
theory, \eg the gauge coupling $g$.

Note that the above criterion guarantees
finiteness of the theory at all orders, although its
conditions involve exclusively one-loop quantities.
The conditions $\b_g^{(1)}=\g^{(1)\ R}_{\ \ \ \ \ S} =0$
are known to guarantee one- and two-loop
vanishing of the $\b$-functions
\cite{parkeswestjonesmezincescu}. Models which
fulfill these conditions are tabulated \eg in
\cite{hamidipateraschwarz}, for the most popular (simple) gauge
groups. Conditions (iii) and (iv) represent
therefore consistency requirements that are necessary
in order to extend the vanishing of
the $\b$-functions at all orders.
To ensure the unicity and non-degeneracy
of the solution of $\g^{(1)\ R}_{\ \ \ \ \ S} =0$ {\it
considered as a solution of} $\b^{(1)}_{rst} = 0$,
one is led to constrain the model by imposing
additional, chiral symmetries. One expects that
such additional symmetries, for some relevant
gauge group, should turn out to have physical
significance and predictive power.

Some models satisfying the all-order finiteness
criterion are known. An $SU(6)$ SYM theory has
been presented in \cite{lucchesipiguetsiboldHPA}.
Other attempts at finding all-order finite
models have resulted in constraining the
initial theory by imposing orbifold symmetries
\cite{zoupanosetc}.

Let me mention that the criterion above can
neither be used for Abelian gauge theories, nor for
semi-simple gauge groups containing $U(1)$ factors.
This is a direct consequence of the form of
$\b^{(1)}_g$ (see \equ{rr0a}). The $U(1)$ quadratic Casimir
being zero, the corresponding
$\b_{g[U(1)]}^{(1)}\neq 0$  and the condition (ii)
of the criterion
cannot be satisfied. This is however physically
fine since one expects a low-energy theory with
a $U(1)$ factor in its gauge group to be an
effective theory, and all-order finiteness to be realized
only above the unification scale.

\section{Conclusions}

In this talk, I have presented a criterion
for all-order vanishing $\b$-functions, \ie perturbative
finiteness, in $N=1$ super-Yang-Mills theories.
Finiteness is expected only at the grand unified
level, since  low-energy, effective theories
are expected to contain a $U(1)$ factor in their gauge
group.

A systematic search for finite SGUTs is
made possible by the fact that the hypotheses of the
criterion involve one-loop quantities only.
Examples of finite $N=1$ SGUTs exist, but no
complete classification has been achieved to date.
{\it The process of testing for all-order
finiteness of a given model is constructive in the
sense that it yields the global symmetries of the
superpotential}. Indeed, one has to look for a unique
and non-degenerate solution of the form
$\l_{\rm Yukawa} =\l_{\rm Yukawa}(g)$ to the
conditions of vanishing one-loop Yukawa
$\b$-functions. Such a solution does generally
not exist, and requires that one restricts the
superpotential until uniqueness and non-degeneracy
are attained.

Finite SGUTs with model-dependent global
(Lie group) or discrete symmetries should provide an
interesting setting for phenomenology. Applying
the criterion presented here could reveal a
precious guide to family symmetry, or to orbifold-type
discrete symmetries resulting from
compactification, as well as to the symmetries of
interest for astrophysics, to mention only a few.


\end{document}